\documentclass[aps,prd,twocolumn,eqsecnum,showpacs,amsmath,nofootinbib]{revtex4}%

\usepackage[dvips]{color,graphicx}
\usepackage{amsfonts,amssymb,theorem,mathrsfs,times}
\newcommand{\ma}[1]{\mbox{$\mathcal{#1}$}}
\newcommand{\qed}{\hbox{\rule[-2pt]{6pt}{6pt}}}
\newcommand{\D}{{\rm d}}

{\theorembodyfont{\upshape}
\newtheorem{Prop}{Proposition}}
{\theorembodyfont{\upshape}
}
{\theorembodyfont{\upshape}
}
{\theorembodyfont{\upshape}
\newtheorem{dn}{Definition}}
{\theorembodyfont{\upshape}
}
{\theorembodyfont{\upshape}
}

\newcommand{\dalm}{\kern1pt\vbox{\hrule height 0.9pt\hbox{\vrule width
0.9pt\hskip 2.5pt\vbox{\vskip 5.5pt}\hskip 3pt\vrule width 0.3pt}\hrule height
0.3pt}\kern1pt}

\begin{document}

\title{
Gauss-Bonnet black holes with non-constant curvature horizons
}

\author{Hideki Maeda}
\email{hideki-at-cecs.cl}


\address{ 
Centro de Estudios Cient\'{\i}ficos (CECS), Casilla 1469, Valdivia, Chile
}

\date{\today}

\begin{abstract} 
We investigate static and dynamical $n(\ge 6)$-dimensional black holes in Einstein-Gauss-Bonnet gravity of which horizons have the isometries of an $(n-2)$-dimensional Einstein space with a condition on its Weyl tensor originally given by Dotti and Gleiser.
Defining a generalized Misner-Sharp quasi-local mass that satisfies the unified first law, we show that most of the properties of the quasi-local mass and the trapping horizon are shared with the case with horizons of constant curvature.
It is shown that the Dotti-Gleiser solution is the unique vacuum solution if the warp factor on the $(n-2)$-dimensional Einstein space is non-constant.
The quasi-local mass becomes constant for the Dotti-Gleiser black hole and satisfies the first law of the black-hole thermodynamics with its Wald entropy.
In the non-negative curvature case with positive Gauss-Bonnet constant and zero cosmological constant, it is shown that the Dotti-Gleiser black hole is thermodynamically unstable.
Even if it becomes locally stable for the non-zero cosmological constant, it cannot be globally stable for the positive cosmological constant.
\end{abstract}



\maketitle

\section{Introduction}
The four-dimensionality of the spacetime is one of the biggest problems in theoretical physics.
One of the possible ways to tackle to this ambitious problem is to set the number of spacetime dimensions as a parameter in order to clarify the characteristic properties of the four-dimensional spacetime.
In this context, Lovelock higher-curvature gravity plays an important role because it is the most natural generalization of general relativity in higher dimensions as the second-order partial differential equations without torsion~\cite{lovelock}.
Among all the classes of Lovelock gravity, the second-order Lovelock gravity, so called the Einstein-Gauss-Bonnet gravity is particularly interesting not only because it is the simplest non-trivial Lovelock gravity but also because it is realized in the low-energy limit of heterotic string theory~\cite{Gross}.

Up to now, not many rigorous results have been established on the generic properties of Lovelock gravity.
Under the present circumstance, the analysis in the symmetric spacetime must be completed to give a firm base for the subsequent study in the less symmetric spacetime.
Hence the $n(\ge 5)$-dimensional spacetime having symmetries corresponding to the isometries of an $(n-2)$-dimensional constant curvature space has been intensively investigated.
The active study of the Gauss-Bonnet black holes was triggered by the discovery of the exact black-hole solution by Boulware and Deser and independently by Wheeler~\cite{GB_BH,Wheeler_1}.
This solution has been generalized to the topological case with a cosmological constant as well as the electric charge~\cite{wiltshire1986,EGBBH,cai}.
Interestingly, the Birkhoff's theorem can be generalized in this system~\cite{wiltshire1986,birkhoff-gb,birkhoff-gb2,mn2008}.
The thermodynamical aspects~\cite{GBBH-thermo} and dynamical stability of the black hole~\cite{GBBH-stability} have been fully investigated and now the Gauss-Bonnet black hole attracts a keen interest from the viewpoint of gauge/gravity correspondence~\cite{GB-ads/cft}. 
The dynamical aspects of the Gauss-Bonnet black holes have been recently studied~\cite{lb2007,mn2008,nm2008,maeda2008}.
(See~\cite{GB-review} for recent reviews.)

One of the characteristic properties of the higher-order Lovelock gravity is that the gravitational equations contain the Riemann tensors explicitly, while the Einstein tensor is written only with the Ricci tensor and the Ricci scalar.
This means that the curvature-effect appears more sharply in the higher-order Lovelock gravity.
This fact is explicitly shown in the solution obtained by Dotti and Gleiser in Einstein-Gauss-Bonnet gravity~\cite{dg2005}.
In general relativity, if we replace the $(n-2)$-dimensional space of positive constant curvature in the Schwarzschild-Tangherlini spacetime by {\it any} $(n-2)$-dimensional Einstein space with positive curvature, the resulting spacetime is still a solution of the vacuum Einstein equations.
However, it is not the case in higher-order Lovelock gravity.
Dotti and Gleiser obtained a $n(\ge 6)$-dimensional black-hole solution of which $(n-2)$-dimensional submanifold is not a constant curvature space but an Einstein space satisfying a certain condition on its Weyl tensor.
The effect of the Weyl tensor appears in the metric function as a parameter and makes the spacetime geometry quite non-trivial.

The Dotti-Gleiser black hole is very important to clarify the nature of the higher-curvature effects. 
The final goal of this paper is to analyze the thermodynamical aspects of the Dotti-Gleiser black hole, which has not been done yet.
Not only because the spacetime has a non-trivial boundary but also because the contribution of the Weyl tensor appears as the slow fall-off in the metric function, we first face the problem of the definition of mass for this black hole.
For this purpose, we take the quasi-local mass approach.
In our recent papers where $(n-2)$-dimensional submanifold is of constant curvature, we showed that the results in general relativity can be generalized in Einstein-Gauss-Bonnet gravity in a unified manner by introducing a well-defined quasi-local mass~\cite{mn2008,nm2008,maeda2008}.
That quasi-local mass is a natural generalization of the Misner-Sharp mass in general relativity, which is considered to be the best one in the spherically symmetric spacetime~\cite{ms1964}.
Remarkably, we will show that a further generalization of this results is possible for the case where the $(n-2)$-dimensional submanifold is an Einstein space satisfying the condition by Dotti and Gleiser.
Reading off the black-hole mass by the generalized Misner-Sharp mass, we study the thermodynamical aspects of the black hole.

The rest of the present paper is constituted as follows.
In the following section, we give the definition of our quasi-local mass and show its basic properties.
Section~III focuses on the study of dynamical black holes defined by a future outer trapping horizon.
The analyses in these two sections are similar to our previous works in~\cite{mn2008,nm2008,maeda2008} and therefore we omit the proofs of several propositions. 
In section IV, we study the properties of the Dotti-Gleiser black hole.
Concluding remarks and discussions including future prospects are summarized in section~V.
In appendix A, we present several examples of the Einstein space satisfying the condition given by Dotti and Gleiser.
In appendix B, the expressions of the curvature tensors are presented.

Our basic notations follow \cite{wald}.
The conventions of curvature tensors are 
$[\nabla _\rho ,\nabla_\sigma]V^\mu ={R^\mu }_{\nu\rho\sigma}V^\nu$ 
and $R_{\mu \nu }={R^\rho }_{\mu \rho \nu }$.
The Minkowski metric is taken to be the mostly plus sign, and 
Roman indices run over all spacetime indices.
We adopt the units in which only the $n$-dimensional gravitational
constant $G_n$ is retained.

\section{Generalized Misner-Sharp mass in Einstein-Gauss-Bonnet gravity}
\subsection{Preliminaries}
The action in the $n (\geq 5)$-dimensional spacetime in Einstein-Gauss-Bonnet gravity 
in the presence of a cosmological constant is given by
\begin{align}
S=&\int\D ^nx\sqrt{-g}\biggl[\frac{1}{2\kappa_n^2}
(R-2\Lambda+\alpha{L}_{\rm GB}) \biggr]+S_{\rm matter},\label{action}\\
{L}_{\rm GB} :=& R^2-4R_{\mu\nu}R^{\mu\nu}
+R_{\mu\nu\rho\sigma}R^{\mu\nu\rho\sigma},
\end{align}
where $\kappa_n := \sqrt{8\pi G_n}$, $G_n$ is the $n$-dimensional gravitational constant, and $L_{\rm GB}$ is called the Gauss-Bonnet term.
$S_{\rm matter}$ in Eq.~(\ref{action}) is the action for matter fields.
$\alpha$ is the coupling constant of the Gauss-Bonnet term. 
This type of action is derived in the low-energy limit 
of heterotic string theory~\cite{Gross} and then $\alpha$ is regarded as the inverse string tension 
and positive-definite. 
In this paper, we don't specify the sign of $\alpha$ unless otherwise noted.
(In our previous works in~\cite{mn2008,nm2008,maeda2008}, $\alpha>0$ was assumed.)

The gravitational equation of the action (\ref{action}) is
\begin{align}
{\cal G}^\mu_{~~\nu}:={G^\mu}_{\nu} +\alpha {H}^\mu_{~~\nu} 
+\Lambda \delta^\mu_{~~\nu}= 
\kappa_n^2 {T}^\mu_{~~\nu}, \label{beq}
\end{align}
where 
\begin{align}
{G}_{\mu\nu}&:= R_{\mu\nu}-{1\over 2}g_{\mu\nu}R,\\
{H}_{\mu\nu}&:= 2\Bigl[RR_{\mu\nu}-2R_{\mu\alpha}
R^\alpha_{~\nu}-2R^{\alpha\beta}R_{\mu\alpha\nu\beta} \nonumber \\
&~~~~~~+R_{\mu}^{~\alpha\beta\gamma}R_{\nu\alpha\beta\gamma}\Bigr]
-{1\over 2}g_{\mu\nu}{L}_{\rm GB}
\end{align}
and ${T}^\mu_{~~\nu}$ is the energy-momentum tensor 
for matter fields obtained from $S_{\rm matter}$.
The field equations (\ref{beq}) contain up to the second derivatives 
of the metric.
In the four-dimensional spacetime, the Gauss-Bonnet term 
does not contribute to the field equations, i.e., $H_{\mu\nu} \equiv 0$, since it
becomes a total derivative.

Suppose the $n$-dimensional spacetime 
$({\ma M}^n, g_{\mu \nu })$ to be a warped product of an 
$(n-2)$-dimensional Einstein space $(K^{n-2}, \gamma _{ij})$
and a two-dimensional orbit spacetime $(M^2, g_{ab})$ under 
the isometries of $(K^{n-2}, \gamma _{ij})$. Namely, the line element
is given by
\begin{align}
g_{\mu \nu }\D x^\mu \D x^\nu =g_{ab}(y)\D y^a\D y^b +r^2(y) \gamma _{ij}(z)
\D z^i\D z^j ,
\label{eq:ansatz}
\end{align} 
where
$a,b = 0, 1;~i,j = 2, ..., n-1$. 
Here $r$ is a scalar on $(M^2, g_{ab})$  with $r=0$ 
defining its boundary and $\gamma_{ij}$ is the metric on $(K^{n-2}, \gamma _{ij})$ with its sectional curvature $k = \pm 1, 0$. 
We assume that $({\ma M}^n, g_{\mu \nu})$ is strongly causal and 
$(K^{n-2}, \gamma _{ij})$ is compact. 
The most general material stress-energy tensor $T_{\mu\nu}$ compatible with this spacetime symmetry is given by
\begin{align}
T_{\mu\nu}\D x^\mu \D x^\nu =T_{ab}(y)\D y^a\D y^b+p(y)r^2 \gamma_{ij}\D z^i\D z^j,
\end{align}  
where $T_{ab}$ and $p$ are a symmetric two-tensor and a scalar on $(M^2, g_{ab})$, respectively.

The $(n-2)$-dimensional Einstein space satisfies
\begin{eqnarray}
\overset{(n-2)}{R}{}_{ijkl}={}^{(n-2)}{C}{}_{ijkl}+k(\gamma_{ik}\gamma_{jl}-\gamma_{il}\gamma_{jk}),
\end{eqnarray}
where ${}^{(n-2)}{C}_{ijkl}$ is the Weyl tensor.
The superscript $(n-2)$ means the geometrical quantity on $(K^{n-2}, \gamma _{ij})$.
If the Weyl tensor is identically zero, $(K^{n-2}, \gamma _{ij})$ is a space of constant curvature.
The Riemann tensor is contracted to give
\begin{eqnarray}
\overset{(n-2)}{R}{}_{ij}&=&k(n-3)\gamma_{ij},\\
\overset{(n-2)}{R}&=&k(n-2)(n-3).
\end{eqnarray}  

In this paper, we consider the Einstein space $(K^{n-2}, \gamma _{ij})$ satisfying a condition
\begin{eqnarray}
\overset{(n-2)}{C}{}^{iklm}\overset{(n-2)}{C}{}_{jklm}=\Theta \delta^i_{~~j}, \label{hc}
\end{eqnarray}  
which is compatible with the Einstein-Gauss-Bonnet equation (2.3) and $\Theta$ is constant.
The condition (\ref{hc}) was introduced originally by Dotti and Gleiser and called the {\it horizon condition}~\cite{dg2005}.
Since $(K^{n-2}, \gamma _{ij})$ is Euclidean, $\Theta$ is non-negative.
The Weyl tensor vanishes identically in three or lower dimensions, so that $\Theta \equiv 0$ holds for $n \le 5$.

Examples of such Einstein spaces with $k \ne 0$ are presented in Appendix A.
Although we don't know any example with $\Theta \ne 0$ and $k=0$, we also consider that case in this paper.
The expressions of the decomposed curvature tensors are given in Appendix B.

\subsection{Generalized Misner-Sharp mass}
We define a scalar function on $(M^2, g_{ab})$ 
with the dimension of mass such that
\begin{align}
\label{qlm}
m_\Theta &:= \frac{(n-2)V_{n-2}^k}{2\kappa_n^2}
\biggl\{-{\tilde \Lambda}r^{n-1}
+r^{n-3}[k-(D r)^2] \nonumber \\
&~~~~~~+{\tilde \alpha}r^{n-5}[k-(Dr)^2]^2+\frac{{\tilde\alpha}{\tilde\Theta}}{n-5}r^{n-5} \biggl\},
\end{align}  
where ${\tilde \alpha} := (n-3)(n-4)\alpha$, ${\tilde \Lambda} := 2\Lambda /[(n-1)(n-2)]$, and ${\tilde\Theta}:=\Theta/[(n-3)(n-4)]$.
$D_a $ is a metric compatible linear connection on $(M^2, g_{ab})$
and $(Dr)^2:=g^{ab}(D_ar)(D_br)$.
$V_{n-2}^k$ denotes the area of $(K^{n-2}, \gamma _{ij})$.
This is a further generalization of the quasi-local mass in the case where $(K^{n-2}, \gamma _{ij})$ is a space of constant curvature, namely $\Theta=0$~\cite{maeda2006b,mn2008}.
In the four-dimensional spherically symmetric case without a cosmological constant, $m_\Theta$ reduces to the Misner-Sharp quasi-local mass~\cite{ms1964}.

The derivative of $m_{\rm \Theta}$ gives
\begin{align}
D_a m_{\Theta} =&\frac{V_{n-2}^kr^{n-2}}{\kappa_n^2}\biggl[{\cal G}_{ab}(D^b r)- {\cal G}^{b}_{b}(D_ar)\biggl],
\end{align}  
where we used the following expressions:
\begin{widetext}
\begin{align}
G_{ab}(D^b r)- G^{b}_{b}(D_ar) =&-(n-2)\frac{D_aD_br}{r}(D^b r)+(n-2)(n-3)\frac{k-(Dr)^2}{2r^2}(D_a r),\\
H_{ab}(D^b r)- H^{b}_{b}(D_ar)=&\frac{2(n-2)(n-3)(n-4)}{r^3}[k-(Dr)^2]\left[(n-5)\frac{[k-(Dr)^2]}{4r}(D_a r)- (D^b r)(D_aD_br) \right] \nonumber \\
&+\frac{(n-2)\Theta}{2r^4}(D_a r).
\end{align}
\end{widetext}
Here the contraction is taken over on the two-dimensional orbit space.
Now it is easy to show that $m_\Theta$ satisfies the unified first law:
\begin{align}
\D  m_\Theta =&A\psi_a\D x^a  +P\D V, \label{1stlaw1} 
\end{align}
where
\begin{align}
\psi ^a :=&{T^a}_bD ^b r +PD ^a r,\\
P:=&-\frac 12 {T^a}_a,\\
A:=&V^k_{n-2}r^{n-2}, \label{area}\\
V:=&\frac{V^k_{n-2}}{n-1}r^{n-1}.
\end{align}
Here $\psi^a$ and $P$ are a vector and a scalar on $(M^2, g_{ab})$, respectively.
$A$ and $V$ are the area and areal volume of the symmetric subspace, respectively, which satisfy $D_a V=AD _ar$.
Equation (\ref{1stlaw1}) does not explicitly contain $\alpha$, $\Lambda$, $k$, and $\Theta$ and has the same form as in four dimensions~\cite{hayward1998}.
In four dimensions, the first and the second terms in the right-hand side are interpreted as the energy flux and the external work, respectively~\cite{hayward1998,ashworth1999}.

We define the locally conserved current vector $J^\mu$ as
\begin{align}
J^\mu :=& -\frac{1}{\kappa _n^2}{\cal G}^\mu_{~~\nu}K^\nu=-{T^\mu}_\nu K^\nu, \label{j} \\
K^\mu  :=&-\epsilon ^{\mu \nu }\nabla _\nu  r. \label{kodamavector}
\end{align}
Here $\epsilon_{\mu \nu}=\epsilon_{ab}(\D x^a)_{\mu}(\D x^b)_{\nu}$, and $\epsilon_{ab}$ is a volume element of $(M^2, g_{ab})$. 
$K^\mu$ is called the Kodama vector~\cite{kodama1980,ms2004}, which is timelike (spacelike) in the untrapped (trapped) region.
The Kodama vector $K^\mu$ is divergent-free and generates a preferred time evolution vector field in the untrapped region.
(This property is discussed in details in~\cite{av2010}.)
Also, $J^\mu$ is divergent-free because of the identity $\nabla_{\nu}{\cal G}^{\mu\nu} \equiv 0$.
In fact, $m_\Theta$ is a quasi-local conserved quantity associated with a locally conserved current $J^\mu$.
This fact is directly observed by another expression of $J^\mu$~\cite{nozawa}:
\begin{align}
J^\mu =&-\frac{1}{V^k_{n-2}}r^{-(n-2)}\biggl(\frac{\partial}{\partial x^a}\biggl)^\mu\epsilon^{ab}D_b m_{\Theta},\nonumber \\
=&-\frac{1}{V^k_{n-2}}r^{-(n-2)}\epsilon^{\mu \nu}\nabla_\nu m_{\Theta},  \label{j-m}
\end{align}
where we used an identity:
\begin{align}
\epsilon^{ab}(D_bD_d r)(D^dr)=&(D^2r)\epsilon^{ad}(D_dr)-( D^aD^br)\epsilon_{bd}(D^dr).
\end{align}
(It is interesting to compare the expressions~(\ref{j}) and (\ref{j-m}) with the results in the stationary case given in section 7 in~\cite{kastor2008}.) 
From Eq.~(\ref{j-m}), we immediately obtain
\begin{align}
\mathscr L_J m_\Theta=&J^\mu \nabla_\mu m_\Theta=0,\label{kodamamass}
\end{align}
which implies that $m_{\Theta}$ is conserved along $J^\mu$.
As a consequence, the integral of $J^\mu $ over some spatial volume with boundary give $m_\Theta$ as an associated charge.
(See section III in~\cite{mn2008}.)

$m_\Theta$ has a monotonic property, which is one of the most important properties for the well-defined quasi-local mass.
In a similar manner to Proposition 4 in~\cite{mn2008}, it is shown that $m_\Theta$ is non-decreasing (non-increasing) in any outgoing (ingoing) spacelike or null direction on an untrapped surface under the dominant energy condition. 

Using the monotonic property, we can show the positivity of $m_{\Theta}$ on the untrapped spacelike hypersurface with a regular center for $\Theta=0$.
(See Proposition 5 in~\cite{mn2008}.)
Positivity is another important property for the well-defined quasi-local mass.
For $\Theta \ne 0$, however, the spacetime cannot be regular at $r=0$ since $k-(Dr)^2=O(r^2)$ in that neighborhood of $r=0$ is not satisfied.

Although it is difficult to show positivity of $m_{\Theta}$ in a generic case, the case with the following special combination between $\Lambda$ and $\alpha$ is exceptional:
\begin{align}
1+4{\tilde\alpha}{\tilde\Lambda}=0, \label{DCgravity}
\end{align}  
which provides a single maximally symmetric vacuum solution for the theory and makes the theory the Chern-Simons gravity in five dimensions~\cite{ctz2000,zanelli2005}.
With the relation~(\ref{DCgravity}), Eq.~(\ref{qlm}) becomes
\begin{align}
m_\Theta =& \frac{(n-2)V_{n-2}^k}{8{\tilde \alpha}\kappa_n^2}r^{n-5}
\biggl\{r^2
+2{\tilde \alpha}[k-(D r)^2]\biggl\}^2 \nonumber \\
&+\frac{(n-2)V_{n-2}^k{\tilde\alpha}{\tilde\Theta}}{2(n-5)\kappa_n^2}r^{n-5},
\end{align}  
and hence $m_{\Theta}>(<) 0$ is satisfied for $\alpha>(<) 0$ without other assumptions.

Solving Eq.~(\ref{qlm}) for $(D r)^2$, we obtain
\begin{align}
\label{trapping}
-(D r)^2 =&-k-\frac{r^2}{2{\tilde\alpha}}\left(1\mp\sqrt{L}\right), \\
L:=&1+\frac{8\kappa_n^2{\tilde\alpha} m_\Theta}{(n-2)V^k_{n-2}r^{n-1}}+4{\tilde\alpha}{\tilde\Lambda}-\frac{4{\tilde\alpha}^2{\tilde\Theta}}{(n-5)r^4}.\label{L}
\end{align}
There are two families of solutions corresponding to 
the sign in front of the square root in Eq.~(\ref{trapping}),
stemming from the quadratic curvature terms in the action.
We call the family having the minus (plus) sign 
the GR-branch (non-GR-branch) solution.
Note that the GR-branch solution has a general
relativistic limit as $\alpha \to 0$, 
\begin{align}
\label{trapping-gr}
-(D r)^2 = 
-k+\frac{2\kappa_n^2m_\Theta}{(n-2)V^k_{n-2}r^{n-3}}+{\tilde\Lambda}r^2,
\end{align}  
but the non-GR branch does not.
Throughout this paper, the upper sign is used for the GR branch.
Also, we assume  
\begin{align}
\label{alphalambda}
1+4{\tilde\alpha}{\tilde\Lambda} \ge 0.
\end{align}  

A branch surface is where two branches of solutions degenerate.
\begin{dn}
\label{def:b-point}
A {\it branch surface} is an $(n-2)$-surface with $L=0$.
\end{dn}
It is shown that the branch surface is generically a curvature singularity.
(The proof is similar to Proposition 12 in~\cite{nm2008}.)
Since $L$ must be non-negative, we obtain $m_\Theta \ge (\le)m_{\rm b}$ for $\alpha>(<)0$, where $m=m_{\rm b}$ is the mass on the branch surface given by
\begin{align}
\label{bc}
m_{\rm b}:=& -\frac{(n-2)(1+4{\tilde\alpha}{\tilde\Lambda})V^k_{n-2}r^{n-1}}
{8\kappa_n^2{\tilde\alpha}} \nonumber \\
&+\frac{(n-2){\tilde\alpha}{\tilde\Theta} V_{n-2}^kr^{n-5}}{2\kappa _n^2(n-5)}.
\end{align}
Hence, $m_\Theta$ has a lower (upper) bound for $\alpha>(<)0$.

\section{Black hole dynamics with non-constant curvature horizons}
In this section, we discuss the dynamical aspects of black holes.
The analysis is performed in the same way as the case with $\Theta=0$ in~\cite{nm2008}, and hence we omit the proofs of several propositions given in this section.

The line element may be written locally in the double-null coordinates as
\begin{align}
\D s^2 = -2e^{-f(u,v)}\D u\D v
+r^2(u,v) \gamma_{ij}\D z^i\D z^j. \label{coords}
\end{align}  
Null vectors $u^\mu(\partial /\partial x^\mu)=(\partial /\partial u)$ and $v^\mu(\partial /\partial x^\mu)=(\partial /\partial v)$ 
are taken to be future-pointing. 
The governing equations~(\ref{beq}) are written as
\begin{widetext}
\begin{align}
&(r_{,uu}+f_{,u}r_{,u})\left[1+\frac{2{\tilde\alpha}}{r^2}
(k+2e^{f}r_{,u}r_{,v})\right]
=-\frac{\kappa_n^2}{n-2} r T_{uu}, \label{equation:uu} \\
&(r_{,vv}+f_{,v}r_{,v})\left[1+\frac{2{\tilde\alpha}}{r^2}
(k+2e^{f}r_{,u}r_{,v})\right]
=-\frac{\kappa_n^2}{n-2} r T_{vv}, \label{equation:vv} \\
&rr_{,uv}+(n-3)r_{,u}r_{,v}+\frac{n-3}{2}k e^{-f}+\frac{{\tilde\alpha}}{2r^2}
[(n-5)k^2e^{-f}+4rr_{,uv}
(k+2e^fr_{,u}r_{,v})+4(n-5)r_{,u}r_{,v}
(k+e^fr_{,u}r_{,v})] \nonumber \\
&~~~~~~-\frac{n-1}{2}{\tilde\Lambda}r^2e^{-f}+\frac{{\tilde\alpha}{\tilde\Theta}}{2r^2}e^{-f}
=\frac{\kappa_n^2}{n-2} r^2T_{uv}, \label{equation:uv} \\
&r^2 f_{,uv}+2(n-3)r_{,u}r_{,v}
+k(n-3)e^{-f}-(n-4)rr_{,uv} \nonumber \\
&~~~~~~+\frac{2{\tilde\alpha}e^{-f}}{r^2}
\biggl[e^f(k+2e^fr_{,u}r_{,v})
\{r^2f_{,uv}-(n-8)rr_{,uv}\}
+2r^2e^{2f}(f_{,u}r_{,u} r_{,vv}
+f_{,v}r_{,v}r_{,uu}) \nonumber \\
&~~~~~~+(n-5)(k+2e^fr_{,u}r_{,v})^2
+2r^2e^{2f}\{r_{,uu}r_{,vv}
+f_{,u}f_{,v}r_{,u}r_{,v}
-(r_{,uv})^2\}\biggl]+\frac{2{\tilde\alpha}{\tilde\Theta}}{r^2}e^{-f} \nonumber \\
&~~~~~~=\kappa_n^2 r^2(T_{uv}+e^{-f}p).
\end{align}  
\end{widetext}

The expansions of two independent future-directed radial null vectors $\partial/\partial v$ and $\partial/\partial u$ are respectively defined as
\begin{align}
\theta_{+}&:=\frac{{\mathscr L}_v A}{A}=(n-2)r^{-1}r_{,v},\\
\theta_{-}&:=\frac{{\mathscr L}_u A}{A}=(n-2)r^{-1}r_{,u},
\end{align}  
where $A$ is the area of the symmetric subspace defined by Eq.~(\ref{area}).
The quasi-local mass $m_\Theta$ is expressed as
\begin{align}
\label{qlm2}
m_\Theta &= \frac{(n-2)V_{n-2}^k}{2\kappa_n^2}r^{n-3}
\biggl[-{\tilde \Lambda}r^2+\left(k+\frac{2}{(n-2)^2} r^2e^{f}
\theta_{+}\theta_{-}\right)\nonumber \\
&~~~~~~
+{\tilde \alpha}r^{-2}\left(k+\frac{2}{(n-2)^2} 
r^2e^{f}\theta_{+}\theta_{-}\right)^2+\frac{{\tilde\alpha}{\tilde\Theta}}{n-5}r^{-2} \biggl]
\end{align}  
and the unified first law (\ref{1stlaw1}) gives
\begin{align}
m_{\Theta,v}&=
\frac{1}{n-2}V_{n-2}^ke^fr^{n-1}(T_{uv}\theta_+-T_{vv}\theta_-), \label{m_v} \\
m_{\Theta,u}&=
\frac{1}{n-2}V_{n-2}^ke^fr^{n-1}(T_{uv}\theta_- -T_{uu}\theta_+). \label{m_u} 
\end{align}  
We impose the energy conditions on the matter field.
The null energy condition for the matter field implies
\begin{align}
T_{uu}\ge 0,~~~T_{vv} \ge 0, \label{nec}
\end{align}
while the dominant energy condition implies
\begin{align}
T_{uu} \ge 0,~~T_{vv}\ge 0,~~T_{uv}\ge 0. \label{dec}
\end{align}  

Now let us study the properties of trapping horizons.
The notion of trapping horizons was originally introduced 
by Hayward~\cite{hayward1994,hayward1996}. 
\begin{dn}
A {\it trapped (untrapped) surface} is a compact $(n-2)$-surface with $\theta_{+}\theta_{-}>(<)0$.
A {\it trapped (untrapped) region} is the union of all trapped (untrapped) surfaces.
A {\it marginal surface} is a compact $(n-2)$-surface with $\theta_{+}\theta_{-}=0$.
\end{dn}
We fix the orientation of the untrapped surface by
$\theta _+>0$ and $\theta_-<0$, i.e., $\partial/\partial u$ and $\partial/\partial v$ are ingoing and outgoing null vectors, respectively.
\begin{dn}
\label{def:4-msphere}
A marginal surface is {\it future} if $\theta_-<0$, {\it past} if
$\theta_{-}>0$, {\it bifurcating} if $\theta_-=0$, {\it outer} if
$\theta_{+,u}<0$, {\it inner} if $\theta_{+,u}>0$ and {\it
degenerate} if $\theta_{+,u}=0$.
\end{dn}
\begin{dn}
\label{def:t-horizon}
A {\it trapping horizon} is the closure of a hypersurface 
foliated by future or past, outer or inner marginal surfaces.
\end{dn}
Among all classes, the {\it future outer} trapping horizon is the most
relevant in the context of black holes~\cite{hayward1994,hayward1996}.
Since the branch surface is generically a curvature singularity, we remove the branch surface from our discussions.

An important fact is that trapping horizons do not always exist.
The following Proposition is shown directly by Eq.~(\ref{trapping}) without using energy condition.
\begin{Prop}
\label{th:absenceTH}
({\it Absence of trapping horizons.}) 
Let $\alpha>0$ be assumed.
Then, an $(n-2)$-surface is necessarily untrapped, and trapping horizons are 
absent in the non-GR-branch solution for $k=0$ and $1$.
In the GR-branch (non-GR-branch) solution for $k=-1$ 
with $r^2<(>)2{\tilde\alpha}$, an $(n-2)$-surface is always trapped 
(untrapped), and trapping horizons are absent.
Next let $\alpha<0$ be assumed.
Then, an $(n-2)$-surface is necessarily trapped, and trapping horizons are 
absent in the non-GR-branch solution for $k=0$ and $-1$.
In the GR-branch (non-GR-branch) solution for $k=1$ 
with $r^2<(>)-2{\tilde\alpha}$, an $(n-2)$-surface is always untrapped 
(trapped), and trapping horizons are absent.
\end{Prop}

\bigskip

By Proposition~\ref{th:absenceTH}, in the case of $\alpha>0$, trapping horizons may exist only for $k=1,0$ and $k=-1$ with $r^2>2{\tilde\alpha}$ in the GR branch and $k=-1$ with $r^2<2{\tilde\alpha}$ in the non-GR branch.
In the case of $\alpha<0$, on the other hand, they may exist only for $k=-1,0$ and $k=1$ with $r^2>-2{\tilde\alpha}$ in the GR branch and $k=1$ with $r^2<-2{\tilde\alpha}$ in the non-GR branch.
In addition to these conditions, an inequality $L(r_{\rm h})>0$ must be satisfied in order to have a horizon in a physical region, where $L$ is given in Eq.~(\ref{L}).

Mass on the trapping horizon $r=r_{\rm h}$is given from Eq.~(\ref{qlm2}) as
\begin{align}
\label{m-h}
m_{\rm h}&:= \frac{(n-2)V_{n-2}^k}{2\kappa_n^2}r_{\rm h}^{n-3}\biggl(-{\tilde \Lambda}r_{\rm h}^2+k+{\tilde \alpha}r_{\rm h}^{-2}k^2+\frac{{\tilde\alpha}{\tilde\Theta}}{n-5}r_{\rm h}^{-2} \biggl).
\end{align}  
Because of the monotonicity of $m_{\Theta}$, the following mass inequality for $m_{\rm h}$ is satisfied~\cite{error}
\begin{Prop}
\label{th:mass}
({\it Mass inequality.}) 
If the dominant energy condition holds, then $m_\Theta \ge  m_{\rm h}(r_{\rm h})$ 
holds on an untrapped spacelike
hypersurface of which the inner boundary is a marginally trapped surface with radius $r_{\rm h}$.
\end{Prop}
For $\alpha \ge 0$ and $\Lambda \le 0$ with $k=1$, Proposition~\ref{th:mass} gives a positive lower-bound for $m_\Theta$ in the black-hole spacetime.

For the trapping horizon, the following propositions are satisfied.
(The proofs are similar to the Propositions 10, 11, and 14 in the case with $\Theta=0$ in~\cite{nm2008}.)
\begin{Prop}
({\it Signature law.})
\label{prop:sig} 
Under the null energy condition, an outer (inner) 
trapping horizon in the GR branch is non-timelike (non-spacelike), 
while it is non-spacelike (non-timelike) in the non-GR branch.
\end{Prop}
\begin{Prop}
({\it Trapped side.})
\label{prop:futureinner}
Let the null energy condition be assumed.
Then, the outside (inside) region of a future inner trapping 
horizon is trapped (untrapped) in the GR
 branch, and the outside (inside) region of a future outer trapping 
horizon is untrapped (trapped) in the non-GR
 branch. 
To the contrary, 
the future (past) domain of a future outer trapping horizon 
is trapped (untrapped) in the GR branch, and the 
future (past) domain of future inner trapping horizon is untrapped
 (trapped) in the non-GR branch.
\end{Prop}
\begin{Prop}
\label{arealaw}
({\it Area law.}) 
Under the null energy condition, the area of a future outer 
(inner) trapping horizon is non-decreasing
 (non-increasing) along the generator of the trapping horizon 
in the GR branch, while it is non-increasing (non-decreasing) 
in the non-GR branch.
\end{Prop}

\bigskip

It is noted that Propositions~\ref{prop:sig}-\ref{arealaw} are satisfied independent of the sign of $\alpha$.
Propositions~\ref{prop:sig} and \ref{prop:futureinner} mean that a future outer trapping horizon in the GR branch is a one-way membrane 
being matched to the concept of a black hole as a region of no escape.
On the other hand, solutions in the non-GR branch have pathological properties.
This is naturally explained by the relation between $T_{\mu \nu }k^\mu k^\nu$ and $R_{\mu \nu }k^\mu k^\nu$ for a radial null vector $k^\mu$, given as
\begin{widetext}
\begin{align}
&\kappa _n^2T_{\mu\nu}k^\mu k^\nu=\pm R_{\mu\nu}k^\mu k^\nu \sqrt{1+\frac{8\kappa _n^2\tilde{\alpha }m_{\Theta}}
{(n-2)V_{n-2}^kr^{n-1}}+4{\tilde\alpha}{\tilde\Lambda}-\frac{4{\tilde\alpha}^2{\tilde\Theta}}{(n-5)r^4}}. \label{Rvv}
\end{align}
\end{widetext}
The null convergence condition $R_{\mu\nu}k^\mu k^\nu \ge 0$ means that gravity is essentially an attractive force.
In general relativity, the null energy condition $T_{\mu\nu}k^\mu k^\nu \ge  0$ ensures the null convergence condition.
It is the same in the GR branch but the null convergence condition is violated in the non-GR branch under the strict null energy condition $T_{\mu\nu}k^\mu k^\nu > 0$.

That's why it was intriguing that the dynamical entropy on the future outer trapping horizon is non-decreasing under the energy condition independent of the branch, shown for $\Theta=0$ with $\alpha>0$~\cite{nm2008}.
Indeed, this entropy increasing law is also satisfied in the case with $\Theta$ independent of the sign of $\alpha$.

Let us first define the dynamical entropy.
The unified first law (\ref{1stlaw1}) can be written as
\begin{widetext}
\begin{align}
A\psi_a=&D_a\biggl[m_\Theta-\frac{(n-2)V_{n-2}^k}{2\kappa_n^2}\biggl\{-{\tilde \Lambda}r^{n-1}+kr^{n-3}+{\tilde \alpha}k^2r^{n-5}+\frac{{\tilde\alpha}{\tilde\Theta}}{n-5}r^{n-5} \biggl\}\biggl] \nonumber \\
&+\frac{(n-2)V_{n-2}^kr^{n-2}}{2\kappa_n^2}\biggl[(n-3)\frac{(Dr)^2}{r^2}+(n-5)\frac{{\tilde\alpha}(Dr)^2\{2k-(Dr)^2\}}{r^4} \biggl]D_a r \nonumber \\
&+\frac{(n-2)V^k_{n-2}r^{n-2}}{2\kappa_n^2}\biggl[\frac{1}{r}+\frac{2{\tilde\alpha}}{r^3}[k-(Dr)^2]\biggl](D^2r) (D_ar).
\end{align}
The first and second terms in the right-hand side vanish on the trapping horizon.
Evaluating on the trapping horizon with its generator $\xi^a$, we obtain  
\begin{align}
A\psi_a\xi^a=&\frac{\kappa_{\rm TH}V_{n-2}^k}{\kappa_n^2}\xi^aD_a \biggl(r^{n-2}+{\tilde \alpha}\frac{2(n-2)}{n-4} kr^{n-4} \biggl),
\end{align}
\end{widetext}
where the dynamical surface gravity $\kappa_{\rm TH}$ is defined by $K^bD_{[b}K_{a]}=\kappa _{\rm TH}K_a$, evaluated on the trapping horizon. 
Here $K^\mu$ is the Kodama vector (\ref{kodamavector}) and we obtain
\begin{align}
\kappa_{\rm TH}:=\frac 12 D^2r\biggl|_{r=r_{\rm h}}.
\end{align}
Identifying the right-hand side as $T_{\rm TH}{\ma L}_\xi S_{\rm TH}$, where $T_{\rm TH}:=\kappa_{\rm TH}/(2\pi)$ is the temperature of the trapping horizon, we obtain the dynamical entropy as
\begin{align}
S_{\rm TH}=&\frac{2\pi V_{n-2}^k}{\kappa_n^2}\biggl(r_{\rm h}^{n-2}+{\tilde \alpha}\frac{2(n-2)}{n-4} kr_{\rm h}^{n-4} \biggl), \nonumber \\
=&\frac{V_{n-2}^kr_{\rm h}^{n-2}}{4G_n}\biggl(1+\frac{2(n-2)(n-3)k\alpha}{r_{\rm h}^2} \biggl),\label{entropy}
\end{align}
where the constant factor was set to zero in order to match the Wald entropy shown in the next section.
It is noted that the entropy formula (\ref{entropy}) does not contain $\Theta$ explicitly.

Also in the case with $\Theta$, the following entropy-increasing law is satisfied independent of the sign of $\alpha$.
(The proofs are similar to the Proposition 15 in the case with $\Theta=0$ in~\cite{nm2008}.)
\begin{Prop}
\label{entropylaw}
({\it Entropy law.}) 
{Under the null energy condition}, the entropy of a future outer 
(inner) trapping horizon is non-decreasing
 (non-increasing) along the generator of the 
trapping horizon in both branches.
\end{Prop}

Up to here, the properties of the trapping horizon were almost the same as the case with $\Theta=0$.
At the end of this section, we show that the positive constant $\Theta$ affects the topology of the trapping horizon.
Evaluation of Eq. (\ref{equation:uv}) on a trapping horizon gives
\begin{align}
&
\frac{(n-2)k}{2r_{\rm h}^2}[n-3+(n-5)\tilde \alpha kr^{-2}_{\rm h}]+\frac{(n-2){\tilde\alpha}{\tilde\Theta}}{2r_{\rm h}^4}
\nonumber \\
&~~~~~
=\kappa _n^2e^fT_{uv}+\Lambda -e^f\theta _{+,u}
\left(1+\frac{2\tilde \alpha k}{r^2_{\rm h}}\right).
\label{Tuveq1}
\end{align}
We notice that $\Theta$ term in the left-hand side gives a positive (negative) contribution for $\alpha>(<)0$.
By Proposition~\ref{th:absenceTH}, the last term in the right-hand side is non-negative for the outer horizon ($\theta _{+,u}<0$) in the GR branch (independent of the sign of $\alpha$).
Hence, the right-hand side of Eq.~(\ref{Tuveq1}) is non-negative for $\Lambda \ge 0$ under the dominant energy condition.
On the other hand, $n-3+(n-5)\tilde \alpha kr^{-2}_{\rm h}>0$ is shown in the GR branch by Proposition~\ref{th:absenceTH}.
Hence, the following proposition holds.
\begin{Prop}
\label{th:topology}
({\it Topology.}) 
Under $\Lambda \ge 0$ and the dominant energy condition, an outer trapping horizon in the GR branch must have a topology of non-negative curvature for $\Theta=0$ and of positive curvature for $\Theta \ne 0$ with $\alpha<0$.
\end{Prop}

\bigskip

In contrary, the horizon topology of negative curvature might be possible in the GR branch for $\Theta \ne 0$ with $\alpha>0$ in the same setting.

\section{Dotti-Gleiser vacuum black hole}
In the last section, we showed that $m_\Theta$ is a natural generalization of the Misner-Sharp mass in the present system.
In this section, we adopt $m_\Theta$ to evaluate the mass of the vacuum black hole.
In the vacuum case, we obtain $m_\Theta=M=$constant because of Eq.~(\ref{1stlaw1}). 
Dotti and Gleiser obtained a static vacuum solution~\cite{dg2005}, of which metric is given as
\begin{align}
\D s^2=-f(r)\D t^2+f^{-1}(r)\D r^2+r^2\gamma_{ij}\D z^i\D z^j,\label{BDW1}
\end{align}
where
\begin{align}
\label{BDW}
f(r) :=& k+\frac{r^2}{2\tilde{\alpha }}\biggl(1\mp \sqrt{h(r)}\biggl), \\
h(r):=&1+\frac{8\kappa _n^2\tilde{\alpha }M}{(n-2)V_{n-2}^kr^{n-1}}+4{\tilde\alpha}{\tilde\Lambda}-\frac{4{\tilde\alpha}^2{\tilde\Theta}}{(n-5)r^4}.
\end{align}
There are curvature singularities at $r=0$ and at $r=r_{\rm b}$ defined by 
\begin{align}
M=& -\frac{(n-2)(1+4{\tilde\alpha}{\tilde\Lambda})V^k_{n-2}r_{\rm b}^{n-1}}
{8\kappa_n^2{\tilde\alpha}} \nonumber \\
&+\frac{(n-2){\tilde\alpha}{\tilde\Theta} V_{n-2}^kr_{\rm b}^{n-5}}{2\kappa _n^2(n-5)}, \nonumber \\
&=:M_{\rm b}(r_{\rm b}).\label{bs-DG}
\end{align}  
The latter is a branch singularity.
For $\Theta=0$, this solution reduces as particular limits to the known solutions obtained before~\cite{GB_BH,Wheeler_1,wiltshire1986,EGBBH,cai}.
$M_{\rm b}$ has a positive (negative) local maximum (mininum) $M_{\rm b}=M_{\rm b(ex)}$ at $r=r_{\rm b(ex)}$ for $\alpha>(<)0$, where $r_{\rm b(ex)}$ and $M_{\rm b(ex)}$ are defined by
\begin{align}
r_{\rm b(ex)}:=&\biggl(\frac{4{\tilde\alpha}^2{\tilde\Theta}}{(n-1)(1+4{\tilde\alpha}{\tilde\Lambda})}\biggl)^{1/4}, \\
M_{\rm b(ex)}:=&\frac{2(n-2){\tilde\alpha}{\tilde\Theta} V_{n-2}^kr_{\rm b(ex)}^{n-5}}{(n-1)(n-5)\kappa _n^2}
\end{align}  
and we have
\begin{align}  
\frac{d^2M_{\rm b}}{dr_{\rm b}^2}\biggl|_{r_{\rm b}=r_{\rm b(ex)}}=& -\frac{4(n-2){\tilde\alpha}{\tilde\Theta} V_{n-2}^k}{2\kappa _n^2}r_{\rm b(ex)}^{n-7}.
\end{align}

\begin{figure}[htbp]
\begin{center}
\includegraphics[width=1.0\linewidth]{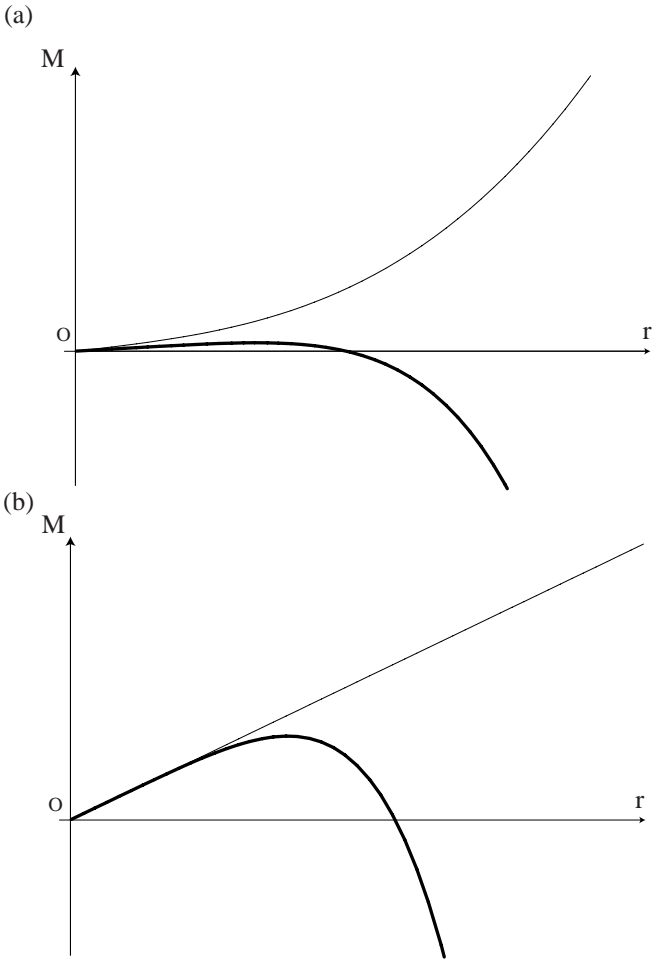}
\caption{\label{fig2} 
The $M=M_{\rm h}(r)$ (\ref{m-hDG}) and $M=M_{\rm b}(r)$ (\ref{bs-DG}) relations for $n=6$, $\alpha>0$, and $\Lambda=0$.
Figures (a) and (b) corresponds to $k=1$ and $k=0$, respectively.
We adopt the unit $(n-2)V_{n-2}^k/(8\kappa_n^2)=1$ and ${\tilde\alpha}=1$.
We put ${\tilde\Theta}=1$ as an example.
(The shape of these curves are qualitatively similar for $n \ge 7$ and $\Theta>0$.)
The thin and thick curves correspond to ${M}={M}_{\rm h}(r)$ and ${M}={M}_{\rm b}(r)$, respectively.
The physical domain of $r$ is given by ${M}>{M}_{\rm b}(r)$.
}
\end{center}
\end{figure}

A Killing horizon $r=r_{\rm h}$ is given by $f(r_{\rm h})=0$.
The relation between the mass and the horizon radius is given by
\begin{align}
M =& \frac{(n-2)V_{n-2}^k}{2\kappa_n^2}\biggl[-{\tilde \Lambda}r_{\rm h}^{n-1}+kr_{\rm h}^{n-3} \nonumber \\
&+{\tilde \alpha}r_{\rm h}^{n-5}\biggl(k^2+\frac{{\tilde\Theta}}{n-5}\biggl) \biggl], \nonumber \\
&=:M_{\rm h}(r_{\rm h}).\label{m-hDG}
\end{align}  
For later convenience, we calculate
\begin{align}
\frac{dM_{\rm h}}{dr_{\rm h}} =& \frac{(n-2)V_{n-2}^k}{2\kappa_n^2}\biggl[-(n-1){\tilde \Lambda}r_{\rm h}^{n-2}+k(n-3)r_{\rm h}^{n-4} \nonumber \\
&+(n-5){\tilde \alpha}r_{\rm h}^{n-6}\biggl(k^2+\frac{{\tilde\Theta}}{n-5}\biggl) \biggl],  \\
\frac{d^2M_{\rm h}}{dr_{\rm h}^2} =& \frac{(n-2)V_{n-2}^k}{2\kappa_n^2}\biggl[-(n-1)(n-2){\tilde \Lambda}r_{\rm h}^{n-3} \nonumber \\
&+k(n-3)(n-4)r_{\rm h}^{n-5} \nonumber \\
&+(n-5)(n-6){\tilde \alpha}r_{\rm h}^{n-7}\biggl(k^2+\frac{{\tilde\Theta}}{n-5}\biggl) \biggl].
\end{align}  
For a certain range of parameters, this Dotti-Gleiser solution admits an outer Killing horizon defined by $df/dr|_{r=r_{\rm h}}>0$, corresponding to the black-hole horizon.
An example is the case with $n \ge 6$, $k=1$, $\alpha>0$, and $\Lambda=0$ in the GR branch.
(See Fig.~\ref{fig2} (a).)
Interestingly, there is also an outer Killing horizon in the case for $n \ge 6$, $k=0$, $\alpha>0$, and $\Lambda=0$ in the GR branch, which is realized only with $\Theta \ne 0$.
(See Fig.~\ref{fig2} (b).)
In these cases, $M=M_{\rm h}(r)$ is monotonically increasing for $r_{\rm h}>0$.
In the presence of a branch singularity for given $M$, we only consider the domain of $r$ connecting to the asymptotic region, namely $r_{\rm b}<r<\infty$.

We obtain
\begin{align}
M_{\rm h}(r)-M_{\rm b}(r)=\frac{(n-2)V_{n-2}^kr^{n-5}(r^2+2{\tilde\alpha}k)^2}{8{\tilde\alpha}\kappa_n^2},
\end{align}
and hence ${M}_{\rm h}\ge {M}_{\rm b}$ is satisfied for $\alpha>0$ with equality holding only at $r=0$ and $r^2=-2{\tilde\alpha}k$ for $k=-1$.
For $\alpha>0$ with $k=1$, the branch singularity is in the untrapped region since $f(r_{\rm b})>0$ is satisfied.
For $n \ge 6$, $k\ge 0$, $\alpha>0$, and $\Lambda=0$ in the GR branch, the asymptotic region $r\to \infty$ is in the untrapped region.
There is a maximum in the curve $M=M_{\rm b}(r)$ at $M=M_{\rm b(ex)}(>0)$.
As a result, there is one non-degenerate outer horizon for $M>M_{\rm b(ex)}$ and no horizon for $M \le M_{\rm b(ex)}$.

\subsection{The Birkhoff's and rigidity theorems}
Here we show that the Dotti-Gleiser solution is the unique vacuum solution with $\Theta \ne 0$ if $r$ is not constant. 
(See~\cite{bcgz2009} for the discussion in a larger class of spacetimes.)
\begin{Prop}
\label{th:vacuum}
({\it Vacuum solutions.})
For $\Theta \ne 0$, an $n$-dimensional vacuum spacetime in Einstein-Gauss-Bonnet
gravity with the metric form (\ref{eq:ansatz}) is isometric to the Dotti-Gleiser solution (\ref{BDW1}) if $r$ is not constant, and the Nariai-type solution if $r$ is constant.
\end{Prop}
\noindent
{\it Proof.}
First we consider the case with $(D r)^2 \ne 0$.
In this case, we can adopt the coordinates such as
\begin{align}
\D s^2=-g(t,r)e^{-2\delta(t,r)}\D t^2+g(t,r)^{-1}\D r^2+r^2\gamma_{ij}\D z^i\D z^j
\end{align}
without loss of generality.
Then the $(t,r)$ or equivalently $(r,t)$ component of the field equation (\ref{beq}) give $g(t,r)=g(r)$ or $g(t,r)=k+r^2/(2{\tilde\alpha})$.
Putting $g(t,r)=k+r^2/(2{\tilde\alpha})$ in the field equation (\ref{beq}), we obtain a single equation
\begin{align}
0=4{\tilde\alpha}^2{\tilde\Theta}-(n-1)(1+4{\tilde\alpha}{\tilde\Lambda})r^4.
\end{align}
This equation is not satisfied for $\Theta \ne 0$ and $\alpha \ne 0$, and hence we obtain $g(t,r)=g(r)$.
Then the equation ${\cal G}^t_{~~t}-{\cal G}^r_{~~r}=0$ gives $g(r)=k+r^2/(2{\tilde\alpha})$ or $\partial\delta(t,r)/\partial r=0$, and hence we obtain $\delta(t,r)={\bar \delta}(t)$.
The function ${\bar \delta}(t)$ can be set to zero by the redefinition of the time coordinate and finally the field equations are integrated to give the Dotti-Gleiser solution, namely $g(r) \equiv f(r)$, 

Next we consider the case with $(D r)^2 =0$.
First we consider the case that $D_a r$ is a null vector.
Then, we can adopt the coordinates such as
\begin{align}
\D s^2=-g(u,r)e^{-2\delta(u,r)}\D u^2-2e^{-\delta(u,r)}\D u\D r+u^2\gamma_{ij}\D z^i\D z^j
\end{align}
without loss of generality.
Then the $(u,u)$ or equivalently $(r,r)$ component of the field equation (\ref{beq}) gives 
\begin{align}
&-2\Lambda u^4+k(n-2)(n-3)u^2 \nonumber \\
&~~~~+(n-2)\alpha[k^2(n-3)(n-4)(n-5)+\Theta]=0.
\end{align}
Since $\Theta$ is positive, the above equation gives a contradiction.
Thus the only possible case for $\Theta \ne 0$ with $(D r)^2 =0$ is that $r$ is a constant $r_0$.
This case was considered in~\cite{md2007}, in which $(M^2,g_{ab})$ is the two-dimensional maximally symmetric spacetime, of which Ricci scalar ${}^{(2)}{R}$ and $r_0$ are given as
\begin{widetext}
\begin{eqnarray}
r_0^2&=&\frac{k(n-2)(n-3)}{4\Lambda}\biggl(1\pm\sqrt{1+\frac{8\alpha\Lambda[k^2(n-3)(n-4)(n-5)+\Theta]}{k^2(n-2)(n-3)^2}}\biggl), \label{Nariai1}\\
{}^{(2)}{R}&=&\frac{2r_0^2}{r_0^2+2\alpha k(n-3)(n-4)}\biggl(\frac{4\Lambda}{n-2}-\frac{k(n-3)}{r_0^2}\biggl). \label{Nariai2}
\end{eqnarray}  
\end{widetext}
This is a Nariai-Bertotti-Robinson-type spacetime~\cite{nbr}, of which quasi-local mass $m_{\rm N}$ is constant given by 
\begin{eqnarray}
m_{\rm N}:= \frac{(n-2)V_{n-2}^k(2{\tilde \Lambda}r_0^2-k)}{(n-5)\kappa_n^2}r_0^{n-3}.
\end{eqnarray}  
The parameters $\Lambda$, $\alpha$, $k$, $n$, and $\Theta$ are taken for $r_0^2$ to be real and positive.
$(M^2,g_{ab})$ is the two-dimensional flat or (A)dS spacetime corresponding to ${}^{(2)}{R}=0$ or ${}^{(2)}{R}(<)>0$.

\qed

\bigskip

Indeed, the vanishing quasi-local mass $m_\Theta = 0$ is related to the vacuum.
The following rigidity theorem holds.
(The proof is similar to Proposition 1 in ~\cite{maeda2008}.)
\begin{Prop}
\label{th:main}
({\it Rigidity.})
Under $\Theta \ne 0$ and the dominant energy condition for $k=1$, $\Lambda\le 0$, and $\alpha >0$, $m_\Theta \equiv 0$ is equivalent to the massless ($M=0$) Dotti-Gleiser solution.
\end{Prop}

\bigskip

Although the Nariai-type solution may have zero quasi-local mass, it does not conflict with the above proposition.
$m_{\rm N}$ certainly becomes zero for the solution (\ref{Nariai1}) with the upper sign and  
\begin{eqnarray}
{\tilde\Theta}= -\frac{k^2(n-5)(1+4{\tilde\alpha}{\tilde \Lambda})}{4{\tilde\alpha}{\tilde \Lambda}}.
\end{eqnarray}  
However, since we have $r_0^2=k/(2{\tilde \Lambda})$ and hence $k\Lambda>0$ is required in this case, the assumption in Proposition~\ref{th:main} is not satisfied.

\subsection{Wald entropy and the first-law of the black-hole thermodynamics}
Finally we discuss the thermodynamical aspects of the Dotti-Gleiser black hole.
Since the quasi-local mass $m_\Theta$ gives $M$ for the Dotti-Gleiser solution (\ref{BDW1}), we identify $M$ as the mass of the Dotti-Gleiser black hole.
As shown below, the mass $M$ and the Wald entropy of the Dotti-Gleiser black hole satisfy the first-law of the black-hole thermodynamics.

Let us calculate the Wald entropy of the Dotti-Gleiser black hole.
The Wald entropy is defined by the following integral performed on $(n-2)$-dimensional spacelike bifurcation surface $\Sigma$~\cite{wald1993,iyerwald1994,jkm1994}:
\begin{align}
S_{\rm W}:=&-2\pi \oint\left(\frac{\partial {\cal L}}{\partial R_{\mu\nu\rho\sigma}}\right)\varepsilon_{\mu\nu}\varepsilon_{\rho\sigma}dV_{n-2}^2,
\end{align}
where $dV_{n-2}^2$ is the volume element on $\Sigma$, $\varepsilon_{\mu\nu}$ is the binormal vector to $\Sigma$ normalized as $\varepsilon_{\mu\nu}\varepsilon^{\mu\nu}=-2$, and ${\cal L}$ is the Lagrangian density.
$\varepsilon_{\mu\nu}$ may be written as $\varepsilon_{\mu\nu}=\xi_{\mu}\eta_{\nu}-\eta_{\mu}\xi_{\nu}$, where $\xi^\mu$ and $\eta^\mu$ are two null vector fields normal to $\Sigma$ satisfying $\xi_\mu\eta^\mu=1$.
The Lagrangian density of Einstein-Gauss-Bonnet gravity is given by 
\begin{align}
{\cal L}=&\frac{1}{2\kappa_n^2}(R-2\Lambda+\alpha{L}_{\rm GB}).
\end{align}
For the later use, we calculate 
\begin{align}
\frac{\partial R}{\partial R_{\mu\nu\rho\sigma}}=&\frac{\partial (g^{\beta\zeta}g^{\alpha\gamma}R_{\alpha\beta\gamma\zeta})}{\partial R_{\mu\nu\rho\sigma}},\nonumber\\
=&g^{\beta\zeta}g^{\alpha\gamma}\delta^{[\mu}_{~[\alpha}\delta^{\nu]}_{~\beta]}\delta^{[\rho}_{~[\gamma}\delta^{\sigma]}_{~\zeta]},\nonumber\\
=&g^{\beta\zeta}g^{\alpha\gamma}\delta^{[\mu}_{~\alpha}\delta^{\nu]}_{~\beta}\delta^{[\rho}_{~\gamma}\delta^{\sigma]}_{~\zeta},\nonumber\\
=&g^{\rho[\mu}g^{\nu]\sigma}.
\end{align}
In a similar manner, we calculate
\begin{align}
\frac{\partial (R_{\alpha\beta}R^{\alpha\beta})}{\partial R_{\mu\nu\rho\sigma}}=&\frac{\partial (g^{\eta\zeta}g^{\gamma\lambda}g^{\alpha\kappa}g^{\beta\omega}R_{\eta\alpha\zeta\beta}R_{\gamma\kappa\lambda\omega})}{\partial R_{\mu\nu\rho\sigma}}, \nonumber \\
=&g^{\eta\zeta}g^{\gamma\lambda}g^{\alpha\kappa}g^{\beta\omega}\delta^{[\mu}_{~[\eta}\delta^{\nu]}_{~\alpha]}\delta^{[\rho}_{~[\zeta}\delta^{\sigma]}_{~\beta]}R_{\gamma\kappa\lambda\omega} \nonumber \\
&+g^{\eta\zeta}g^{\gamma\lambda}g^{\alpha\kappa}g^{\beta\omega}\delta^{[\mu}_{~[\gamma}\delta^{\nu]}_{~\kappa]}\delta^{[\rho}_{~[\lambda}\delta^{\sigma]}_{~\omega]}R_{\eta\alpha\zeta\beta},\nonumber \\
=&2g^{\eta\zeta}g^{\alpha\kappa}g^{\beta\omega}\delta^{[\mu}_{~\eta}\delta^{\nu]}_{~\alpha}\delta^{[\rho}_{~\zeta}\delta^{\sigma]}_{~\beta}R_{\kappa\omega} \nonumber \\
=&2g^{\kappa[\nu}g^{\mu][\rho}g^{\sigma]\omega}R_{\kappa\omega}
\end{align}
and
\begin{align}
\frac{\partial (R_{\alpha\beta\gamma\eta}R^{\alpha\beta\gamma\eta})}{\partial R_{\mu\nu\rho\sigma}}=&\frac{\partial (g^{\alpha\zeta}g^{\beta\lambda}g^{\gamma\kappa}g^{\eta\omega}R_{\alpha\beta\gamma\eta}R_{\zeta\lambda\kappa\omega})}{\partial R_{\mu\nu\rho\sigma}},\nonumber \\
=&g^{\alpha\zeta}g^{\beta\lambda}g^{\gamma\kappa}g^{\eta\omega}\delta^{[\mu}_{~[\alpha}\delta^{\nu]}_{~\beta]}\delta^{[\rho}_{~[\gamma}\delta^{\sigma]}_{~\eta]}R_{\zeta\lambda\kappa\omega} \nonumber \\
&+g^{\alpha\zeta}g^{\beta\lambda}g^{\gamma\kappa}g^{\eta\omega}\delta^{[\mu}_{~[\zeta}\delta^{\nu]}_{~\lambda]}\delta^{[\rho}_{~[\kappa}\delta^{\omega]}_{~\zeta]}R_{\alpha\beta\gamma\eta},\nonumber \\
=&2\delta^{[\mu}_{~\alpha}\delta^{\nu]}_{~\beta}\delta^{[\rho}_{~\gamma}\delta^{\sigma]}_{~\eta}R^{\alpha\beta\gamma\eta},\nonumber \\
=&2R^{\mu\nu\rho\sigma}.
\end{align}

For the metric (\ref{BDW1}), $\Sigma$ is given by $t=$constant and $r=r_{\rm h}=$constant and we have $\varepsilon_{tr}=1$.
Then, we obtain
\begin{align}
\frac{\partial R}{\partial R_{\mu\nu\rho\sigma}}\varepsilon_{\mu\nu}\varepsilon_{\rho\sigma}=&g^{\rho[\mu}g^{\nu]\sigma}\varepsilon_{\mu\nu}\varepsilon_{\rho\sigma}, \nonumber \\
=&2g^{rr}g^{tt}\varepsilon_{tr}\varepsilon_{tr},\nonumber \\
=&-2
\end{align}
and
\begin{align}
\frac{\partial L_{\rm GB}}{\partial R_{\mu\nu\rho\sigma}}\varepsilon_{\mu\nu}\varepsilon_{\rho\sigma}=&\varepsilon_{\mu\nu}\varepsilon_{\rho\sigma}\biggl[2Rg^{\rho[\mu}g^{\nu]\sigma} \nonumber \\
&-8g^{\kappa[\nu}g^{\mu][\rho}g^{\sigma]\omega}R_{\kappa\omega}+2R^{\mu\nu\rho\sigma}\biggl], \nonumber \\
=&-4R+8g^{rr}R_{rr}+8g^{tt}R_{tt}+8R^{trtr}, \nonumber \\
=&-\frac{4(n-2)(n-3)(k-f(r))}{r^2}.
\end{align}
Finally, the Wald entropy is given by
\begin{align}
S_{\rm W}=&-\frac{2\pi}{16\pi G_n} \oint\left(\frac{\partial (R+\alpha L_{\rm GB}))}{\partial R_{abcd}}\right)\varepsilon_{ab}\varepsilon_{cd}r^{n-2}d\Omega_{n-2}^2,\nonumber \\
=&-\frac{1}{8 G_n} \biggl[-2-\frac{4(n-2)(n-3)k\alpha}{r_{\rm h}^2}\biggl]r_{\rm h}^{n-2}\int d\Omega_{n-2}^2,\nonumber \\
=&\frac{A_{\rm h}}{4 G_n} \biggl[1+\frac{2(n-2)(n-3)k\alpha}{r_{\rm h}^2}\biggl],
\end{align}
where $A_{\rm h}$ is the area of the horizon.
This coincides with the form of the dynamical entropy (\ref{entropy}).

We show that the first-law of the black-hole thermodynamics is satisfied for the Dotti-Gleiser black hole.
The variation of Eq.~(\ref{m-hDG}) with respect to $r_{\rm h}$ is 
\begin{align}
\delta M =& \frac{(n-2)V_{n-2}^k}{2\kappa_n^2}\biggl[-(n-1){\tilde \Lambda}r_{\rm h}^{n-2}+(n-3)kr_{\rm h}^{n-4} \nonumber \\
&+(n-5){\tilde \alpha}r_{\rm h}^{n-6}\biggl(k^2+\frac{{\tilde\Theta}}{n-5}\biggl) \biggl]\delta r_{\rm h}.\label{id2}
\end{align}  
On the other hand, the variation of $S_{\rm W}$ with respect to $r_{\rm h}$ is 
\begin{align}
\delta S_{\rm W}=& \frac{2\pi(n-2)V_{n-2}^k}{\kappa_n^2}r_{\rm h}^{n-5}(r_{\rm h}^2+2k{\tilde \alpha})\delta r_{\rm h}.\label{id3}
\end{align}  
The surface gravity $\kappa:=(1/2)(df/dr)|_{r=r_{\rm h}}$ is given by
\begin{align}
\kappa=& \frac{-(n-1){\tilde \Lambda}r_{\rm h}^4+(n-3)kr_{\rm h}^2+(n-5)k^2{\tilde \alpha}+{\tilde \alpha}{\tilde\Theta}}{2r_{\rm h}(r_{\rm h}^2+2k{\tilde \alpha})}.\label{kappa}
\end{align}  
From Eq.~(\ref{id2}), (\ref{id3}), and (\ref{kappa}), we obtain
\begin{align}
\delta M = T\delta S_{\rm W},
\end{align}  
where the temperature of the horizon defined by $T:=\kappa/(2\pi)$.
This is the first-law of the black-hole thermodynamics.

\subsection{Thermodynamical stability}
Subsequently, we discuss the thermodynamical stability of the Dotti-Gleiser black holes.
The heat capacity $C$ is given by
\begin{align}
C:=\frac{dM}{dT}=\frac{dM}{dr_{\rm h}}\biggl/\frac{dT}{dr_{\rm h}}.
\end{align}
The positive heat capacity means the local thermodynamical stability.
$dT/dr_{\rm h}$ and $dM/dr_{\rm h}$ are given as
\begin{align}
\frac{dT}{dr_{\rm h}}=&\frac{B}{4\pi r_{\rm h}^2(r_{\rm h}^2+2k{\tilde \alpha})^2}, \label{dTdr} \\
B:=&-(n-1){\tilde \Lambda}r_{\rm h}^4(r_{\rm h}^2+6k{\tilde \alpha}) \nonumber \\
&-(n-3)k\biggl(r_{\rm h}^2+\frac{n-9}{2(n-3)}k{\tilde \alpha}\biggl)^2  \nonumber \\
&-\frac{7n^2-46n+39}{4(n-3)}k^3{\tilde \alpha}^2-3{\tilde \alpha}{\tilde\Theta}\biggl(r_{\rm h}^2+\frac23k{\tilde \alpha}\biggl),\\
\frac{dM}{dr_{\rm h}}=&\frac{(n-2)V_{n-2}^k}{2\kappa_n^2}\biggl[-(n-1){\tilde \Lambda}r_{\rm h}^{n-2}+(n-3)kr_{\rm h}^{n-4} \nonumber \\
&~~~~~~~~~~~~~~~+(n-5){\tilde \alpha}r_{\rm h}^{n-6}\biggl(k^2+\frac{{\tilde\Theta}}{n-5}\biggl) \biggl].\label{dMdr}
\end{align}  

If the system is thermodynamically locally stable, we may discuss the global thermodynamical stability by the free energy defined by $F:=M-TS$.
If $F<0$, the system is thermodynamically globally stable.
The free energy of the Dotti-Gleiser black hole is given as
\begin{align}
F=&\frac{V_{n-2}^kr_{\rm h}^{n-5}D}{2\kappa_n^2(n-4)(n-5)(r_{\rm h}^{2}+2k{\tilde \alpha})}, \label{F}\\
D:=&(n-4)(n-5){\tilde \Lambda}r_{\rm h}^6+6(n-2)(n-5)k{\tilde \alpha}{\tilde \Lambda}r_{\rm h}^4 \nonumber \\
&+(n-4)(n-5)k\biggl(r_{\rm h}^2+\frac{n-8}{2(n-4)}{\tilde\alpha}k\biggl)^2 \nonumber \\
&+\frac{n(n-5)(7n-32)}{4(n-4)}{\tilde\alpha}^2k^3 \nonumber \\
&+{\tilde\alpha}{\tilde\Theta}[3(n-4)r_{\rm h}^2+2(n-2)k{\tilde\alpha}].
\end{align}  

The thermodynamical stability depends on the parameters $\Lambda$, $\alpha$, $k$, $\Theta$, and $n$ and therefore it is quite complicated to clarify the parameter-dependence of the thermodynamical stability.
(See~\cite{cai} for the analysis with $\Theta=0$.)
In addition, the Dotti-Gleiser black hole may have a branch singularity at a finite physical radius $r=r_{\rm b}$ which makes the allowed position of the horizon not being $0<r_{\rm h}<\infty$.
Nevertheless, the thermal instability is proven analytically for some particular cases.
\begin{Prop}
\label{th:stability}
({\it Thermodynamical instability.})
Suppose $\alpha>0$ and $\Theta\ne 0$ (and hence $n\ge 6$).
Then, $C<0$ is satisfied for $r_{\rm h}>0$ in the case with $\Lambda=0$ and $k \ge 0$.
Also, $F>0$ is satisfied for $r_{\rm h}>0$ in the case with $\Lambda \ge 0$ and $k \ge 0$.
\end{Prop}
\noindent
{\it Proof}. 
Let us consider the case with $\alpha>0$, $\Theta\ne 0$, and $k \ge 0$.
Since $7n^2-46n+39>0$ for $n>5$, we show $dT/dr_{\rm h}<0$ by Eq.~(\ref{dTdr}) for any $r_{\rm h}>0$ in the case of $\Lambda \ge 0$.
Also, we show $dM/dr_{\rm h}>0$ by Eq.~(\ref{dMdr}) for any $r_{\rm h}>0$ in the case of $\Lambda \le 0$.
Hence, $C<0$ is satisfied for $\Lambda=0$.
It is also seen in Eq.~(\ref{F}) that $F>0$ is satisfied for any $r_{\rm h}>0$ in the case of $\Lambda \ge 0$.
\qed

\bigskip

We already showed the existence of the black-hole configuration for $k \ge 0$, $\alpha>0$, $\Lambda=0$, and $\Theta\ne 0$ at the beginning of this section.
(See Fig.~\ref{fig2}.)
Proposition~\ref{th:stability} implies that the Dotti-Gleiser black hole cannot be thermodynamically locally stable in such a case.
Even if it becomes locally stable for $\Lambda \ne 0$, $\alpha > 0$, and $k \ge 0$, it cannot be globally stable for $\Lambda>0$.

Closing this section, we discuss the formation and the evaporation of the Dotti-Gleiser black hole.
In the case without the Weyl term, the Boulware-Deser-Wheeler black hole can be formed from the gravitational collapse of some matter cloud with a regular center~\cite{maeda2006b,GB-collapse}.
In contrary, it is not straightforward to imagine such a classical formation process in the presence of the Weyl term because it violates the regularity condition at the center.
Nevertheless, the Dotti-Gleiser black hole could be formed by some quantum process.
After the formation, the black hole loses its mass and shrinks by the Hawking radiation.
Let us consider the case of $\Lambda=0$, $\alpha>0$, and $k=1$, for example.
In this case, there is a positive lower bound $M=M_{\rm b(ex)}(>0)$ for the black-hole mass. (See Fig.~\ref{fig2} (a).)
Since the temperature of the black hole is still non-zero at $M=M_{\rm b(ex)}$, a non-central naked singularity suddenly appears after some moment as a final state of the Hawking radiation.
This evolution scenario is characteristic in the presence of the Weyl term.

\section{Summary}
\label{sec:summary}

In this paper, we studied the properties of static and dynamical black holes in the symmetric spacetime in Einstein-Gauss-Bonnet gravity.
We assumed that the $n$-dimensional spacetime is a cross product of the two-dimensional Lorentzian spacetime and a $(n-2)$-dimensional Einstein space with a condition on the Weyl tensor given by Dotti and Gleiser.
We constructed an infinite sequence of such Einstein spaces as cross products of many Einstein spaces satisfying that condition.

Although the effect of the Weyl tensor appears non-trivially in the field equations, the unified first law is shown to hold by introducing a natural generalization of the Misner-Sharp quasi-local mass.
It was shown that that quasi-local mass has the monotonic property and most of the dynamical properties of black holes are shared with the case where the $(n-2)$-dimensional submanifold is of constant curvature.

In the vacuum case, we showed the Birkhoff's theorem stating that the Dotti-Gleiser solution is the unique vacuum solution in the case where the warp factor of the $(n-2)$-dimensional submanifold is non-constant.
The mass and the Wald entropy of the Dotti-Gleiser black hole satisfy the first-law of the black-hole thermodynamics.
We derived the heat capacity and the free energy of the Dotti-Gleiser black hole and showed that it cannot be thermodynamically locally stable for $\Lambda=0$, $\alpha > 0$, $\Theta\ne 0$, and $k \ge 0$.
Even if it becomes locally stable for $\Lambda \ne 0$, $\alpha > 0$, $\Theta\ne 0$, and $k \ge 0$, it cannot be globally stable for $\Lambda>0$.

As a future study, it is important to clarify the parameter-dependence of the thermodynamical stability of the Dotti-Gleiser black hole.
The analysis seems rather complicated because of the many parameters as well as the branch singularity at a finite radius.
For the comprehensive study in a systematic way, the approach using the $M$-$r_{\rm h}$ diagram presented in~\cite{tm2005} will be useful.


\section*{Acknowledgments}
The author thanks Masato Nozawa, Sourya Ray, and Julio Oliva for many helpful comments. 
This work was funded by the Fondecyt grants 1100328. 
The Centro de Estudios Cient\'{\i}ficos (CECS) is funded by
the Chilean Government through the Millennium Science Initiative and
the Centers of Excellence Base Financing Program of Conicyt, and Conicyt grant "Southern
Theoretical Physics Laboratory" ACT-91. 
CECS is also supported by a group of private companies which at present
includes Antofagasta Minerals, Arauco, Empresas CMPC, Indura,
Naviera Ultragas, and Telef\'{o}nica del Sur.

\appendix

\section{Einstein spaces satisfying the horizon condition}
\label{sec:summary}
In~\cite{dg2005}, Dotti and Gleiser gave a unique example with $k=1$ satisfying the horizon condition (\ref{hc}) in the class of the Bohm metric~\cite{bohm1998,ghp2003}.
It is a cross-product of two $m(\ge 3)$-dimensional spheres. This example is contained in the following proposition 11 as a special case.
Another non-trivial example satisfying the horizon condition was given in~\cite{bcgz2009}, which is a class of the Bergman space.
The metric is
\begin{align}
\gamma_{ij}dx^idx^j=&\frac{4}{p(1-{\bar l}^2\rho^2)^2}\biggl[\frac{d\rho^2}{1+{\bar l}^2\rho^2} \nonumber \\
&+\rho^2(1-{\bar l}^2\rho^2)(\sigma_1^2+\sigma_2^2)+\rho^2(1+{\bar l}^2\rho^2)\sigma_3^2\biggl], \label{bergman}\\
\sigma_1:=&\frac12(\sin\theta\cos\psi\D\phi-\sin\psi\D\theta),\\
\sigma_2:=&\frac12(\sin\theta\sin\psi\D\phi+\cos\psi\D\theta),\\
\sigma_3:=&\frac12(\cos\theta\D\phi+\D\psi),
\end{align}  
where ${\bar l}$ and $p$ are constant, with which we obtain
\begin{align}
{}^{(4)}R^i_{~j}=3p{\bar l}^2\delta^i_{~j},\quad \Theta=6p^2{\bar l}^4.
\end{align}  
$\sigma_i~(i=1,2,3)$ are the standard bases of left invariant one-forms on $SU(2)$.
In order to satisfy ${}^{(d)}R_{ij}=k(d-1)\gamma_{ij}$ with $d=4$ and $k=1$, we set $p{\bar l}^2=1$ and hence we have $\Theta=6$.

In~\cite{bcgz2009}, it was also shown that a four-dimensional Einstein space $S^2\times S^2$ satisfies the horizon condition.
Its metric and curvature quantities are 
\begin{align}
\gamma_{ij}dx^idx^j=&\rho_1^2(d\Omega_2^2+d\Omega_2^2),\\
{}^{(4)}R^i_{~j}=&\frac{1}{\rho_1^2}\delta^i_{~j},\quad \Theta=\frac{4}{3\rho_1^4}.
\end{align}  
In order to satisfy ${}^{(d)}R_{ij}=k(d-1)\gamma_{ij}$ with $d=4$ and $k=1$, we set $\rho_1^2=1/3$ and hence we have $\Theta=12$.

In fact, if ${K}^p$ ($p\ge 2$) is an Einstein space satisfying the horizon condition, then the $d(=p\times q)$-dimensional product space ${\ma M}^d \approx \underset{q}{\underbrace{{K}^p\times \cdots\times {K}^p}}$ ($q\ge 2$) is again an Einstein space satisfying the horizon condition.
\begin{Prop}
\label{th:product}
({\it Einstein spaces satisfying the horizon condition.})
If ${K}^p (p\ge 2)$ is an Einstein space satisfying the horizon condition (containing the case of $\Theta=0$), then the $d(=p\times q)$-dimensional product space ${\ma M}^d \approx \underset{q}{\underbrace{{K}^p\times \cdots\times {K}^p}} (q\ge 2)$ with the same warp factors also satisfies the horizon condition.
\end{Prop}
\noindent
{\it Proof}. 
We consider a $d(=p\times q)$-dimensional product space $({\ma M}^d,\gamma_{ij})$ of which metric is given as
\begin{eqnarray}
\gamma_{ij}=\mbox{diag}(r_0^2{\bar \gamma}_{a_1b_1},\cdots,r_0^2{\bar \gamma}_{a_qb_q}),
\label{eq:structure}
\end{eqnarray}
where $i,j=1,2, \cdots, d$ and $a_\sigma, b_\sigma=(\sigma-1)q+1,(\sigma-1)q+2,\cdots,\sigma q$ for $\sigma =1,\cdots,q$.
$r_0$ is a constant and ${\bar \gamma}_{a_\sigma b_\sigma }$ is the metric on ${K}^{p}$ satisfying the horizon condition, namely
\begin{align}
\overset{(p)}{C}_{{a_\sigma }{b_\sigma }{d_\sigma }{e_\sigma }}\overset{(p)}{C}{}^{{f_\sigma }{b_\sigma }{d_\sigma }{e_\sigma }}={\bar \Theta}\delta^{{f_\sigma }}_{{a_\sigma }}
\end{align}  
is satisfied for each $\sigma $, where ${\bar \Theta}$ is a non-negative constant and the superscript $(p)$ means that it is a geometric quantity on $({K}^p,{\bar \gamma}_{a_\sigma b_\sigma })$.

A $p$-dimensional Einstein space $({K}^p,{\bar\gamma}_{a_\sigma b_\sigma })$ satisfies
\begin{align}
\overset{(p)}{R}{}_{a_\sigma b_\sigma d_\sigma e_\sigma }=\overset{(p)}{C}{}_{a_\sigma b_\sigma d_\sigma e_\sigma }+{\bar k}({\bar\gamma}_{a_\sigma d_\sigma }{\bar\gamma}_{b_\sigma e_\sigma }-{\bar\gamma}_{a_\sigma e_\sigma }{\bar\gamma}_{b_\sigma d_\sigma }),
\end{align}
which is contracted to give
\begin{eqnarray}
\overset{(p)}{R}{}_{a_\sigma b_\sigma }&=&{\bar k}(p-1){\bar\gamma}_{a_\sigma b_\sigma },\\
\overset{(p)}{R}&=&{\bar k}p(p-1).
\end{eqnarray}  
The metric ${\bar \gamma}_{a_\sigma b_\sigma }$ may be chosen such that the curvature constant ${\bar k}$ becomes ${\bar k}=\pm 1,0$.

Non-zero components of the Riemann tensor, Ricci tensor, and the Ricci scalar on $({\ma M}^d,\gamma_{ij})$ are
\begin{align}
\overset{(d)}{R}{}^{a_\sigma }_{~~{b_\sigma }{d_\sigma }{e_\sigma }}=&\overset{(p)}{R}{}^{a_\sigma }_{~~{b_\sigma }{d_\sigma }{e_\sigma }}, \nonumber \\
=&{\bar k}(\delta^{a_\sigma }_{d_\sigma }{\bar \gamma}_{{b_\sigma }{e_\sigma }}-\delta^{a_\sigma }_{e_\sigma }{\bar \gamma}_{{b_\sigma }{d_\sigma }})+\overset{(p)}{C}{}^{a_\sigma }_{~~{b_\sigma }{d_\sigma }{e_\sigma }},\\
\overset{(d)}{R}{}_{{b_\sigma }{e_\sigma }}=&\sum_{k}\overset{(d)}{R}{}^{a_k}_{~~{b_\sigma }{a_k}{e_\sigma }}=\overset{(d)}{R}{}^{a_\sigma }_{~~{b_\sigma }{a_\sigma }{e_\sigma }}=\overset{(p)}{R}{}^{a_\sigma }_{~~{b_\sigma }{a_\sigma }{e_\sigma }}, \nonumber \\
=&(p-1){\bar k}{\bar \gamma}_{{b_\sigma }{e_\sigma }},\label{Ricci-d}\\
\overset{(d)}{R}=&\gamma^{ij}\overset{(d)}{R}{}_{ij}=r_0^{-2}\sum_{k}{\bar \gamma}^{{b_k}{e_k}}\overset{(d)}{R}{}_{{b_k}{e_k}} ,\nonumber \\
=&r_0^{-2}(p-1){\bar k}\sum_{k}{\bar \gamma}^{{b_k}{e_k}}{\bar \gamma}_{{b_k}{e_k}}, \nonumber \\
=&r_0^{-2}pq(p-1){\bar k}.
\end{align} 
Here the superscript $(d)$ means that it is a geometric quantity on $({\ma M}^d,\gamma_{ij})$.
Equation~(\ref{Ricci-d}) implies that $({\ma M}^d,\gamma_{ij})$ is also an Einstein space satisfying ${}^{(d)}R_{ij}=r_0^{-2}(p-1){\bar k}{\gamma}_{ij}$.
By choosing $r_0^2$ to be 
\begin{eqnarray}
r_0^2=\frac{(p-1)|{\bar k}|}{d-1} \label{normal}
\end{eqnarray}  
for ${\bar k}\ne 0$, the curvature constant $k$ on $({\ma M}^d,\gamma_{ij})$ is set to be $k=\pm 1$.

The Weyl tensor on $({\ma M}^d,\gamma_{ij})$ is written in terms of the Riemann tensor as
\begin{widetext}
\begin{align}
\overset{(d)}{C}_{{a_\sigma }{b_\rho }{d_\kappa }{e_\omega }}=&\overset{(d)}{R}_{{a_\sigma }{b_\rho }{d_\kappa }{e_\omega }}-\frac{2}{d-2}(\gamma_{{a_\sigma }[{d_\kappa }}\overset{(d)}{R}_{{e_\omega }]{b_\rho }}-\gamma_{b_\rho [{d_\kappa }}\overset{(d)}{R}_{{e_\omega }]{a_\sigma }})+\frac{2}{(d-1)(d-2)}\overset{(d)}{R}\gamma_{{a_\sigma }[{d_\kappa }}\gamma_{e_\omega ]{b_\rho }}.
\end{align}  
Let us consider which set of $(\rho,\kappa,\omega)$ gives a non-zero component for a given $\sigma$.
The component for $\sigma=\kappa\ne \rho=\omega$ is 
\begin{align}
\overset{(d)}{C}_{{a_\sigma }{b_\rho }{d_\sigma }{e_\rho }}=&-\frac{1}{d-2}(\gamma_{{a_\sigma }{d_\sigma }}\overset{(d)}{R}_{{b_\rho }{e_\rho }}+\gamma_{{b_\rho }{e_\rho }}\overset{(d)}{R}_{{d_\sigma }{a_\sigma }})+\frac{1}{(d-1)(d-2)}(\overset{(d)}{R}\gamma_{{a_\sigma }{d_\sigma }}\gamma_{{e_\rho }{b_\rho }}), \nonumber \\
=&-\frac{2r_0^2(p-1){\bar k}}{d-2}{\bar \gamma}_{{a_\sigma }{d_\sigma }}{\bar \gamma}_{{b_\rho }{e_\rho }}+\frac{r_0^2pq(p-1){\bar k}}{(d-1)(d-2)}{\bar \gamma}_{{a_\sigma }{d_\sigma }}{\bar \gamma}_{{e_\rho }{b_\rho }}, \nonumber \\
=&-\frac{(p-1)r_0^2{\bar k}}{d-1}{\bar \gamma}_{{a_\sigma }{d_\sigma }}{\bar \gamma}_{{b_\rho }{e_\rho }}.
\end{align}  
The component for $\sigma=\omega\ne \kappa=\rho$ is
\begin{align}
\overset{(d)}{C}_{{a_\sigma }{b_\kappa }{d_\kappa }{e_\sigma }}=&\frac{(p-1)r_0^2{\bar k}}{d-1}{\bar \gamma}_{{a_\sigma }{e_\sigma }}{\bar \gamma}_{{b_\kappa }{d_\kappa }}.
\end{align}  
The component for $\sigma=\kappa=\rho=\omega$ is 
\begin{align}
\overset{(d)}{C}_{{a_\sigma }{b_\sigma }{d_\sigma }{e_\sigma }}=&\overset{(d)}{R}_{{a_\sigma }{b_\sigma }{d_\sigma }{e_\sigma }}-\frac{1}{d-2}(\gamma_{{a_\sigma }{d_\sigma }}\overset{(d)}{R}_{{b_\sigma }{e_\sigma }}-\gamma_{{a_\sigma }{e_\sigma }}\overset{(d)}{R}_{{d_\sigma }{b_\sigma }}-\gamma_{{d_\sigma }{b_\sigma }}\overset{(d)}{R}_{{e_\sigma }{a_\sigma }}+\gamma_{{b_\sigma }{e_\sigma }}\overset{(d)}{R}_{{d_\sigma }{a_\sigma }}) \nonumber \\
&+\frac{1}{(d-1)(d-2)}(\overset{(d)}{R}\gamma_{{a_\sigma }{d_\sigma }}\gamma_{{e_\sigma }{b_\sigma }}-\overset{(d)}{R}\gamma_{{a_\sigma }{e_\sigma }}\gamma_{{d_\sigma }{b_\sigma }}), \nonumber \\
=&r_0^2[{\bar k}({\bar \gamma}_{{a_\sigma }d_\sigma }{\bar \gamma}_{{b_\sigma }{e_\sigma }}-{\bar \gamma}_{{a_\sigma }e_\sigma }{\bar \gamma}_{{b_\sigma }{d_\sigma }})+\overset{(p)}{C}{}_{{a_\sigma }{b_\sigma }{d_\sigma }{e_\sigma }}]-\frac{2(p-1){\bar k}r_0^2}{d-2}({\bar \gamma}_{{a_\sigma }{d_\sigma }}{\bar \gamma}_{{b_\sigma }{e_\sigma }}-{\bar \gamma}_{{a_\sigma }{e_\sigma }}{\bar \gamma}_{{d_\sigma }{b_\sigma }}) \nonumber \\
&+\frac{qp(p-1){\bar k}r_0^2}{(d-1)(d-2)}({\bar \gamma}_{{a_\sigma }{d_\sigma }}{\bar \gamma}_{{e_\sigma }{b_\sigma }}-{\bar \gamma}_{{a_\sigma }{e_\sigma }}{\bar \gamma}_{{d_\sigma }{b_\sigma }}), \nonumber \\
=&r_0^2\biggl[\overset{(p)}{C}{}_{{a_\sigma }{b_\sigma }{d_\sigma }{e_\sigma }}+\frac{(d-p){\bar k}}{d-1}({\bar \gamma}_{{a_\sigma }{d_\sigma }}{\bar \gamma}_{{e_\sigma }{b_\sigma }}-{\bar \gamma}_{{a_\sigma }{e_\sigma }}{\bar \gamma}_{{d_\sigma }{b_\sigma }})\biggl].
\end{align}  
Other sets of $(\rho,\kappa,\omega)$ are zero.
Hence, we obtain
\begin{align}
\overset{(d)}{C}_{{a_\sigma }ijk}\overset{(d)}{C}{}^{{f_\eta }ijk}=&\sum_{\rho,\kappa,\omega}\overset{(d)}{C}_{{a_\sigma }{b_\rho }{d_\kappa }{e_\omega }}\overset{(d)}{C}{}^{{f_\eta }{b_\rho }{d_\kappa }{e_\omega }}, \nonumber \\
=&\overset{(d)}{C}_{{a_\sigma }{b_\sigma }{d_\sigma }{e_\sigma }}\overset{(d)}{C}{}^{{f_\eta }{b_\sigma }{d_\sigma }{e_\sigma }}+\sum_{\rho\ne \sigma}\overset{(d)}{C}_{{a_\sigma }{b_\rho }{d_\sigma }{e_\rho }}\overset{(d)}{C}{}^{{f_\eta }{b_\rho }{d_\sigma }{e_\rho }}+\sum_{\rho\ne \sigma}\overset{(d)}{C}_{{a_\sigma }{b_\rho }{d_\rho }{e_\sigma }}\overset{(d)}{C}{}^{{f_\eta }{b_\rho }{d_\rho }{e_\sigma }}, \nonumber \\
=&\overset{(d)}{C}_{{a_\sigma }{b_\sigma }{d_\sigma }{e_\sigma }}\overset{(d)}{C}{}^{{f_\eta }{b_\sigma }{d_\sigma }{e_\sigma }}+\sum_{\rho\ne \sigma}\frac{(p-1)^2r_0^{-4}{\bar k}^2}{(d-1)^2}{\bar \gamma}_{{a_\sigma }{d_\sigma }}{\bar \gamma}_{{e_\rho }{b_\rho }}{\bar \gamma}^{{f_\eta }{d_\sigma }}{\bar \gamma}^{{e_\rho }{b_\rho }} \nonumber \\
&~~~~~~~~~~~~~~~+\sum_{\rho\ne \sigma}\frac{(p-1)^2r_0^{-4}{\bar k}^2}{(d-1)^2}{\bar \gamma}_{{a_\sigma }{e_\sigma }}{\bar \gamma}_{{d_\rho }{b_\rho }}{\bar \gamma}^{{f_\eta }{e_\sigma }}{\bar \gamma}^{{d_\rho }{b_\rho }}, \nonumber \\
=&\overset{(d)}{C}_{{a_\sigma }{b_\sigma }{d_\sigma }{e_\sigma }}\overset{(d)}{C}{}^{{f_\eta }{b_\sigma }{d_\sigma }{e_\sigma }}+\frac{p(q-1)(p-1)^2r_0^{-4}{\bar k}^2}{(d-1)^2}\delta^{{f_\eta }}_{{a_\sigma }}+\frac{p(q-1)(p-1)^2r_0^{-4}{\bar k}^2}{(d-1)^2}\delta^{{f_\eta }}_{{a_\sigma }}, \nonumber \\
=&r_0^{-4}\biggl[\overset{(p)}{C}{}_{{a_\sigma }{b_\sigma }{d_\sigma }{e_\sigma }}+\frac{(d-p){\bar k}}{d-1}({\bar \gamma}_{{a_\sigma }{d_\sigma }}{\bar \gamma}_{{e_\sigma }{b_\sigma }}-{\bar \gamma}_{{a_\sigma }{e_\sigma }}{\bar \gamma}_{{d_\sigma }{b_\sigma }})\biggl] \nonumber \\
&\times \biggl[\overset{(p)}{C}{}^{{f_\eta }{b_\sigma }{d_\sigma }{e_\sigma }}+\frac{(d-p){\bar k}}{d-1}({\bar \gamma}^{{f_\eta }{d_\sigma }}{\bar \gamma}^{{e_\sigma }{b_\sigma }}-{\bar \gamma}^{{f_\eta }{e_\sigma }}{\bar \gamma}^{{d_\sigma }{b_\sigma }})\biggl]+\frac{2p(q-1)(p-1)^2r_0^{-4}{\bar k}^2}{(d-1)^2}\delta^{{f_\eta }}_{{a_\sigma }}, \nonumber \\
=&r_0^{-4}\biggl[{\bar \Theta}+\frac{2(p-1)(d-p)^2{\bar k}^2}{(d-1)^2}+\frac{2p(q-1)(p-1)^2{\bar k}^2}{(d-1)^2}\biggl]\delta^{{f_\eta }}_{{a_\sigma }}, \nonumber \\
=&r_0^{-4}\biggl[{\bar \Theta}+\frac{2p(p-1)(q-1){\bar k}^2}{(d-1)}\biggl]\delta^{{f_\eta }}_{{a_\sigma }}.
\end{align}
\end{widetext}
Writing the horizon condition on $({\ma M}^d,\gamma_{ij})$ as
\begin{align}
\overset{(d)}{C}_{jklm}\overset{(d)}{C}{}^{iklm}=\Theta\delta^{i}_{j},
\end{align}  
we finally obtain the relation between $\Theta$ and ${\bar \Theta}$ as
\begin{align}
\Theta=r_0^{-4}\biggl[{\bar \Theta}+\frac{2p(p-1)(q-1){\bar k}^2}{(d-1)}\biggl].
\end{align}

\qed

\section{Tensor decomposition}
In this appendix, we present the decompositions of the Einstein and Gauss-Bonnet tensors.
The non-vanishing components of the Levi-Civit\'a connections are
\begin{align}
\begin{aligned}
{\Gamma ^a}_{bc}&=\overset{(2)}{\Gamma}{}^a_{bc }(y),\quad 
{\Gamma ^i}_{ij}={\hat{\Gamma} ^i}_{~jk}(z), \\
{\Gamma ^a}_{ij}&=-r (D^a r) \gamma _{ij},\quad 
{\Gamma ^i}_{ja}=\frac{D_a r}{r}{\delta ^i}_j, 
\end{aligned}
\end{align}
where the superscript (2) denotes the two-dimensional quantity,
and $D_a$ is the two-dimensional linear connection compatible with
$g_{ab}$. ${\hat \Gamma ^i}_{~jk}$ is the Levi-Civit\'a connection
associated with $\gamma _{ij}$.
The Riemann tensor is given by
\begin{align}
{R^a}_{bcd}&=\overset{(2)}{R}{}^a_{~~bcd},\nonumber \\
{R^a}_{ibj}&=-r(D^a D_b r)\gamma _{ij},
\label{eq:Riemann}\\
{R^i}_{jkl}&=[k-(Dr)^2]({\delta ^i}_k\gamma _{jl}
-{\delta ^i}_l\gamma _{jk})+\overset{(n-2)}{C}{}^i_{~~jkl}. \nonumber 
\end{align}
The Ricci tensor and the Ricci scalar are given by 
\begin{align}
R_{ab}&=\overset{(2)}{R}_{ab}-(n-2)\frac{D_aD_br}{r}, \nonumber \\
R_{ij}&=\left\{-r D^2r+(n-3)[k-(Dr)^2]\right\}
\gamma _{ij}, \label{eq:Ricci} \\
R&=\overset{(2)}{R}-2(n-2)\frac{D^2r}{r}
+(n-2)(n-3)\frac{k-(Dr)^2}{r^2}. \nonumber
\end{align}
\begin{widetext}
The Einstein tensor is given by
\begin{align}
G_{ab}=&-(n-2)\frac{D_aD_br}{r}+g_{ab}\biggl[(n-2)\frac{D^2r}{r}-(n-2)(n-3)\frac{k-(Dr)^2}{2r^2}\biggl], \\
G_{ij}=&\biggl[-\frac12 r^2\overset{(2)}{R}+ (n-3)rD^2r-\frac12 (n-3)(n-4)[k-(Dr)^2]\biggl]\gamma _{ij},
\end{align}
while the Gauss-Bonnet tensor is
\begin{align}
H_{ab}=&\frac{2(n-2)(n-3)(n-4)}{r^3}[k-(Dr)^2]\left[\left\{D^2r-(n-5)\frac{[k-(Dr)^2]}{4r}\right\}g_{ab}- D_aD_br \right]-\frac{(n-2)\Theta}{2r^4}g_{ab}, \\
H_{ij}=& 2(n-3)(n-4)\left[-\frac {k-(Dr)^2}2{}^{(2)}R- (D^2r)^2 +(D_aD_br)(D^aD^br)\right. \nonumber 
\\
&\left. \qquad \qquad \qquad \quad -(n-5)(n-6)\frac{[k-(Dr)^2]^2}{4 r^2}+(n-5)\frac{k-(Dr)^2}{r}D^2r\right]\gamma _{ij}-\frac{(n-6)\Theta}{2r^2}\gamma_{ij}.
\end{align}
The following combinations are useful:
\begin{align}
G_{ab}-G^d_{~~d}g_{ab}=&-(n-2)\frac{D_aD_br}{r}+(n-2)(n-3)\frac{k-(Dr)^2}{2r^2}g_{ab},\\
H_{ab}-H^d_{~~d}g_{ab}=&\frac{2(n-2)(n-3)(n-4)}{r^3}[k-(Dr)^2]\left[- D_aD_br-(D^2r)g_{ab} +(n-5)\frac{[k-(Dr)^2]}{4r}g_{ab}\right]+\frac{(n-2)\Theta}{2r^4}g_{ab},
\end{align}
where the contraction is taken over on the two-dimensional orbit space.
The Gauss-Bonnet Lagrangian is given by
\begin{align}
L_{\rm GB}=\frac{4 (n-2) (n-3)}{r^2} 
\biggl[&\frac{k-(Dr)^2}{2 }{}^{(2)}R+(D^2 r)^2-{(D_a D_b r) (D^a D^b r)} \nonumber \\
&+(n-4)(n-5)\frac{[k-(D r)^2]^2}{4 r^2}-(n-4)
\frac{k-(Dr)^2}{r}D^2 r\biggl]+\frac{(n-2)\Theta}{2r^4}.
\end{align}
\end{widetext}


\end{document}